# The effect of solidification direction with respect to gravity on ice-templated TiO$_2$ microstructures

*Kristen L. Scotti[1], Lauren G. Kearney[1], Jared Burns[1], Matthew Ocana[1], Lucas Duros[2], Aaron Shelhamer[3], and David C. Dunand[1*]*

[1]Department of Materials Science and Engineering, Northwestern University, Evanston, IL 60208, USA
[2]Department of Mechanical Engineering, Trinity College, Hartford, CT 06106, USA
[3]Department of Computer Science, Northern Illinois University, Dekalb, IL 60112, USA

**Abstract**
Unidirectional ice-templating produces materials with aligned, elongated pores via: (i) directional solidification of particle suspensions wherein suspended particles are rejected and incorporated between aligned dendrites, (ii) sublimation of the solidified fluid, and (iii) sintering of the particles into elongated walls which are templated by the ice dendrites. Most ice-templating studies utilize upward solidification techniques, where solid ice is located at the bottom of the solidification mold (closest to the cold source), the liquid suspension is on top of the ice, and the solidification front advances upward, against gravity. Liquid water reaches its maximum density at 4°C; thus, liquid nearest the cold source is less dense than warmer liquid above (up to 4°C, above which, a density inversion occurs, and liquid density decreases with increasing temperature). The lower density liquid nearest the cold source is expected to rise due to buoyancy, promoting convective fluid motion during solidification. Here, we investigate the effect of solidification direction with respect to the direction of gravity on ice-templated microstructures to study the role of buoyancy-driven fluid motion during solidification. We hypothesize that, for upward solidification, the convective fluid motion that results from a liquid density gradient occurs near the solidification front. For downward solidification, we expect that this fluid motion occurs farther away from the solidification front. Aqueous suspensions of TiO$_2$ nanoparticles (10-30 nm in size, 10, 15, and 21 vol.%) are solidified upward (against gravity, with ice on bottom and water on top), downward (water on bottom, ice on top), and horizontally (perpendicular to gravity). Microstructural investigation of sintered samples shows evidence of buoyancy-driven, convective fluid flow during solidification for samples solidified upwards (against gravity), including (i) tilting of the wall (and pore) orientation with respect to the induced temperature gradient, (ii) ice lens defects (cracks oriented parallel to the freezing direction), and (iii) radial macrosegregation. These features are not observed for downward nor horizontal solidification configurations, consistent with the hypothesis that convective fluid motion does not interact directly with the solidification front for downward solidification.

---

[*] Corresponding author. Tel.: +1 847 491 5370. Fax: +1 847 467 6573
E-mail address: dunand@northwestern.edu (D.C. Dunand)

# 1. Introduction

Unidirectional ice-templating has been utilized extensively since the early 2000's for the fabrication of materials that exhibit aligned, anisotropic pore structures [1-6]. In general, the process consists of three steps. First, an aqueous suspension of particles is solidified in the presence of a thermal gradient. The solidification front, which consists of colonies of ice dendrites, propagates unidirectionally. Suspended particles are first rejected by advancing dendrites and later incorporated within interdendritic space. After solidification, ice is removed via sublimation. The resulting green body consists of elongated, particle-packed walls separated by pore channels that nearly replicate ice dendrite morphology. Finally, the green body is sintered to densify the particle-packed walls. The versatility of the technique is well demonstrated: ice-templating has been shown to be a viable method for the production of ceramic [7, 8], metallic [9-11], polymeric [12, 13], pharmaceutical [14], and foodstuff [15] materials. When non-aqueous fluids are used, the processing method is more generally referred to as "freeze-casting." Camphene and tert-butyl alcohol are the most commonly utilized non-aqueous fluids; nevertheless, aqueous freeze-casting is used in the vast majority of reported studies [5].

Liquid water reaches its maximum density at 4°C (~1.0000 g·cm$^{-3}$) and decreases with both increasing and decreasing temperature: at -2 and 10 °C, the density is ~0.9997 g·cm$^{-3}$ [16]. Thus, a monotonic temperature profile imposed during directional solidification creates a non-monotonic density gradient in the water. The direction and magnitude of the density gradient depend on: (i) the direction of solidification and (ii) the temperatures of the liquid at the cold versus hot side of the suspension, respectively. For upward solidification (solid ice on bottom; liquid suspension on top; **Fig. 1**-a), the solidification front advances against gravity and the liquid closest to the cold source is less dense than the warmer liquid above and unstable with respect to buoyancy; the cold water (0-4ºC) rises. In alloy solidification, buoyancy-driven flow has been shown to result in microstructural inhomogeneities through mechanisms such as dendrite fragmentation, freckling, and the "upstream effect" (dendrites tilt in the direction of flow) [4-6]. For ice-templated, sintered yttria-stabilized zirconia, Bettge *et al*. [17] observed cigar-shaped pores intermixed with typical lamellar channels in cross-sectional images that were taken perpendicular to the solidification direction. The authors posited that the cigar-shaped pore structures resulted from Rayleigh-Bénard convective cells during solidification, which they attributed to the density inversion of water.

Unidirectional ice-templating suspensions are almost always solidified vertically upward, against Earth's gravity vector [18-24]. The reason for this convention is not addressed in the literature, but it is likely attributable to ease of set-up. A few cases can be found where horizontal [25-27] and downward [28-31] solidification techniques were utilized. Of these studies, only Groβberger *et al.* [29] acknowledge that the freezing direction may influence microstructures templated during the solidification process. These researchers fabricated ice-templated alumina using a vacuum-induced surface freezing technique. Aqueous suspensions containing 3-13 vol.% $Al_2O_3$ (500 nm) were held under vacuum and solidification was initiated at the top surface of the suspension; the solidification front proceeded downward, along the gravity vector. The authors reported that ice-lens defects, which are sometimes observed in ice-templated materials solidified vertically upward [32-34], were not observed in their materials and posited that their absence could be due to differing gravitational effects during downward (vs. upward) solidification.

Ice lens defects arise during solidification as ice-filled platelets that grow perpendicular to the freezing direction (unlike dendritic lamellae, which grow parallel to the freezing direction).

After sublimation and sintering, ice lenses result in cracks which are oriented perpendicular to the freezing direction [33-37]. The formation of ice lenses during solidification of aqueous particle suspensions has been attributed to a pattern of particle engulfment, rather than rejection, at the solidification interface [2, 37-41]. During directional solidification, particles are pushed ahead of the solidification front prior to their incorporation within interdendritic space; this creates an enriched particle region (or "accumulation layer") that is located immediately ahead of the solid/liquid interface. During upward solidification, particle sedimentation acts to further increase the particle fraction within the accumulation region. If the particle fraction at the interface exceeds a so-called "breakthrough concentration" [42], ice dendrites are unable to propagate through the particle-enriched region. The particles within the accumulation region are engulfed by the solid/liquid interface and a particle-free ice layer forms immediately thereafter; this process may repeat provided water and particles are available for incorporation in the ice/particle composite. For downward solidification, particles sediment away from the solidification interface; thus, sedimentation acts to deplete the particle fraction within the accumulation region.

Groβberger *et al.* [29] calculated the sedimentation velocity of their suspended, submicron particles (~500 nm $Al_2O_3$) as 0.3 $\mu m \cdot s^{-1}$ and determined that, for upward solidification, the sedimentation velocity would have been too slow relative to the solidification velocity (30 $\mu m \cdot s^{-1}$) to cause an appreciable build-up of particles at the interface. On this basis, the authors argued that gravitational effects were too weak to contribute to ice lens development during solidification of submicron particle suspensions; thus, a different factor must be responsible for their absence in materials fabricated via the downward solidification, vacuum-induced surface freezing technique.

However, upward solidification control experiments were not carried out and other gravity-driven effects (*e.g.,* buoyancy of the liquid) were not considered.

In a previous study using parabolic flights [43], we investigated the effect of gravity during upward solidification of aqueous suspensions (5-20 wt.%) of nanometric $TiO_2$ by comparing sintered microstructures of materials solidified under normal terrestrial gravity (1 $g$) to those solidified under reduced gravity (*i.e.*, Martian, lunar and micro-gravity, corresponding to 0.38, 0.16 and ~0$g$). Ice lens defects were observed in samples solidified from the most concentrated, 20 wt.% $TiO_2$ suspensions that were solidified upward under 1$g$, but not under any of the reduced gravity conditions. The particle size employed (10-30 nm) was much smaller than those employed for the alumina materials described by Groβberger *et al.* [29] (500 nm). Additional microstructural differences were observed, including improved homogeneity of wall orientation in samples solidified under reduced gravity. In all cases, solidification proceeded vertically upward and the microstructural differences between samples solidified under reduced and terrestrial gravity were attributed to gravity-driven convective fluid flow (*i.e.*, liquid buoyancy) during solidification.

Here, we investigate the effect of solidification direction, under constant terrestrial gravity (1$g$), on ice-templated microstructures to better understand the role of buoyancy-driven fluid motion during solidification. Aqueous suspensions of $TiO_2$ nanoparticles (10-30 nm, the same as for our previous study [43]) are solidified upward (opposite gravity; solid on bottom, liquid on top), downward (with gravity, liquid on bottom, solid on top), and horizontally (perpendicular to gravity; **Fig. 1**). Microstructural investigation of sintered samples provides evidence that buoyancy-driven convective fluid flow induces microstructural inhomogeneities in samples solidified upward; no such evidence is observed for downward nor horizontal solidification directions. This is consistent with the hypothesis that, during downward (and possibly horizontal) solidification, buoyancy-driven fluid motion occurs far enough away from the interface that the disruption of dendritic growth at the interface is minimized. To the best of our knowledge, this is the first study that systematically investigates the effect of solidification direction on microstructures created *via* the ice templating technique.

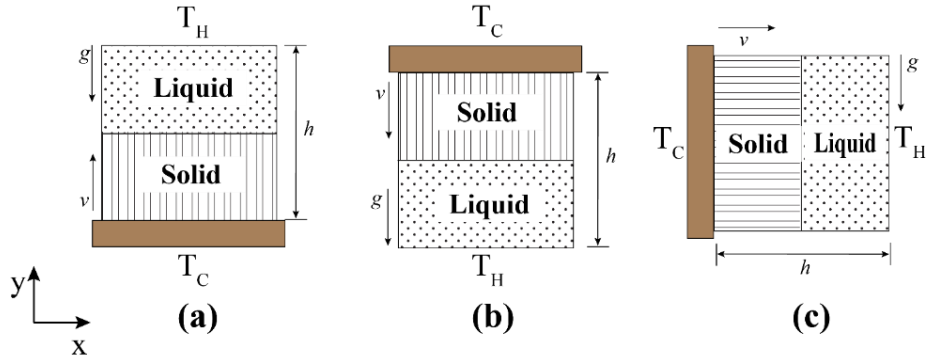

**Fig. 1.** Schematic of experimental set-up for (a) "Up", (b) "Down", and (c) "Horizontal," solidification conditions where $T_H$ and $T_C$ represent the hot and cold side of the mold, respectively, and h is the mold height which is filled completely with the $TiO_2$ aqueous suspension. The solidification direction ($v$) is (a) upward, (b) downward, and (c) horizontal to the direction of gravity ($g$).

## 2. Experimental methods
### 2.1. Suspension preparation

Suspensions of titanium dioxide ($TiO_2$) nanoparticles in deionized water were prepared using a mixture of ethylene glycol (Consolidated Chemical & Solvents, Quakertown, PA) and ammonium hydroxide (SEOH, Navasota, TX) as dispersants, and agar (NOW Foods, Bloomingdale, IL) as a binder. Agar (0.2 wt.% with respect to $TiO_2$) was added to deionized water which was boiled to dissolve the agar and degas the solution. Ethylene glycol (5 vol.%, with respect to water) and three weight fractions (30, 40, and 50 wt.% with respect to liquid, corresponding to volume fractions of 10, 15, and 21%) of $TiO_2$ nanoparticles (anatase phase, 99.5% purity, 10-30 nm, specific surface area 50 $m^2g^{-1}$, SkySpring Nanomaterials, Houston, TX) were added to the liquid solution. Ammonium hydroxide was then added drop-wise to obtain a suspension pH of 10 and suspensions were stirred for 30 minutes using a magnetic stirrer. The colloidal suspensions (held at ambient temperature) were injected into solidification molds consisting of a PVC tubes (15 mm inner diameter, 1.6 mm wall thickness). A cylindrical aluminum base (1.9 mm thickness) sealed one side of the PVC tubes and a rubber stopper (25 mm thickness) sealed the opposite side. The rubber stoppers contained two holes: suspensions were injected into one hole until fluid leaked out of the second hole. Both holes were subsequently closed off using epoxy. The suspension fill height was kept to ~3 mm to allow for comparison to previous samples solidified on parabolic flights (wherein sample height was constrained due to microgravity time durations of ~25 s) [44].

## 2.2. Ice-templating, sublimation, and sintering

The top surface of a copper box containing dry ice (solid $CO_2$ sublimating at 195 K) was utilized as a freezing substrate. The substrate temperature, recorded throughout solidification using a J-type thermocouple, was 228±4 K. Samples were solidified under three conditions with respect to the direction of solidification front propagation: (i) upward (liquid on top, solid on bottom), (ii) downward (solid on top, liquid on bottom), and (iii) horizontally to the gravity vector. These solidification directions are depicted in **Fig. 1** and are referred to as "Up," "Down,", and "Horizontal," respectively. In all cases, a rubber-elastic belt (held at approximately the same tension) was used to hold the solidification mold in contact with the copper box and to ensure consistent thermal contact for all samples, irrespective of solidification direction. Solidification times were estimated by visual inspection (noting the time when the last-to-solidify regions of samples appeared to be frozen through the transparent PVC mold) at ~35-45 s; samples were left on the copper box for ~2 additional minutes to ensure suspensions were fully solidified prior to removal.

Solidified samples were sublimated in a freeze-dryer (Labcono, Freeze Dry System) for at least 24 h at 233 K and low residual pressure (< 3 Pa). After sublimation, samples were placed on an alumina plate and sintered in a box furnace at 1173 K for 1 h in air, using heating and cooling rates of 5 K·min$^{-1}$.

**2.3. Microstructural characterization**

Ceramographic examination was conducted using scanning electron and optical microscopy on sintered samples which were epoxy-mounted, ground and polished. Pore and wall measurements were obtained from optical micrographs. For these investigations, longitudinal cross-sections (with the center of each image corresponds roughly to the center of the respective sintered sample, with approximate dimensions of 15 x 3 mm) of polished samples were scanned using a Nikon MA200 automated microscope, which was set to obtain 15x15 pixel-grid images across the width and height of each cross-section; a 20% overlap per pixel-grid image was used. Pixel grids were subsequently stitched to obtain full cross-section images. Stitched images were divided equally both radially (x3) and vertically (x3), for a total of nine images per cross-section, for further analysis (see supplementary materials, Fig. S1). Pore walls were segmented on resulting images using ImageJ/Fiji [45] and applying the Otsu threshold algorithm [46] on contrast-normalized images [47]. A total of 90 samples were analyzed: ten samples from each $TiO_2$ weight fraction (30, 40, and 50%) for each solidification direction ("up", "down", and "horizontal").

Pore width and wall width were measured over the width and height of each binary image using an in-house program written in C++ and employing the ImageMagick, Magick++ [48] library to decode images into raw pixel color data. For each pixel row in an image, the number of consecutive pixels of each color (white or black) are counted as a segment and the width of each segment (number of pixels) is added to a "black" or "white" vector container. After all segments in each row are counted, a csv file is generated, which provides the number of white or black pixels per segment for each row in the image. For each segment, the number of pixels was converted to a length using the scale (pixels/µm) of the corresponding image. Approximately eight million measurements of each pore and wall widths were obtained for the 90 samples analyzed.

Scanning electron microscope (SEM) images were obtained using a Hitatchi S-3400N-VP SEM operating at an accelerating voltage of 20 kV; these images were used to make qualitative assessments of pore and wall morphology and for measuring secondary dendritic arms. Due to the low electrical conductivity of $TiO_2$, backscatter detection under low vacuum was utilized.

Python was used for all statistical analyses. Data are expressed as mean ± standard deviation. Paired $t$-tests were used to compare means of two groups; one-way ANOVA and post hoc, paired $t$-tests with Bonferroni corrections were used to compare means between more than two groups. A probability value of $p<0.01$ was utilized to determine statistical significance.

## 3. Results and Discussion
### 3.1. Solidification, porosity, and shrinkage

The temperature of the freezing substrate, recorded throughout solidification using a J-type thermocouple, was 228±4 K. No statistically significant differences were detected between freezing substrate temperatures for temperatures acquired when samples were placed on the substrate to those for which they were taken off. Similarly, the freezing substrate temperature did not vary significantly based on solidification direction. Solidification times were estimated by visual inspection at ~35-45 s; thus, average solidification velocity for the ~3 mm sample height was 75±10 µm·s$^{-1}$. Horizontal and downward solidification requires suspension filling procedures that minimize the transfer of air pockets. Whereas air bubbles rise harmlessly to the suspension surface during upward solidification, they collect at the solid/liquid interface during downward and horizontal solidification. An SEM image showing the microstructural consequence of this bubble/solidification front interaction during horizontal solidification is shown in Fig. S-2. No microstructural data were obtained from samples containing bubble-defects and those samples are not included in the sample count shown in

**Table 1**.

Macroscopic sintering shrinkage, measured as the change in diameter after solidification to that measured after sintering, was 25±5% without statistically significant differences present in terms of $TiO_2$ weight fraction nor solidification direction (Table 1). Macroscopic porosity, measured by image analysis, was observed to decrease with increasing weight fractions of $TiO_2$ (51±10, 48±7, and 42±9% for 30, 40, and 50 wt.% $TiO_2$, respectively;

**Table 1**). This is consistent with previous investigations [23, 49] and expected because macroporosity obtained after sublimation and sintering is dependent on the total volume fraction of ice, and thus of fluid in the initial suspension. No significant porosity differences were found among the three solidification directions for any given $TiO_2$ weight fraction.

**Table 1**. Summary of microstructure parameters (pore and wall width, and porosity) based on optical ceramographic investigation of 90 samples ($N$ = number of samples) created from aqueous suspensions with various TiO$_2$ weight fractions solidified against ("Up"), with ("Down"), and perpendicular to ("Horizontal") the gravity vector.

| TiO$_2$ (wt.%) | Direction | $N$ | Pore width (µm) | | | Wall width (µm) | | | Porosity (%) |
|---|---|---|---|---|---|---|---|---|---|
| | | | Overall | Center | Radial | Overall | Center | Radial | |
| 30 (10 vol.%) | Up | 10 | 27±22 | 39±27 | 19±12 | 32±22 | 56±37 | 30±22 | 51±10 |
| | Down | 10 | 7±5 | 6±4 | 6±4 | 8±6 | 8±5 | 9±6 | |
| | Horizontal | 10 | 10±6 | 11±6 | 8±4 | 14±7 | 14±7 | 14±7 | |
| 40 (15 vol.%) | Up | 10 | 27±18 | 30±18 | 23±13 | 30±16 | 36±19 | 29±14 | 48±7 |
| | Down | 10 | 17±13 | 16±9 | 17±10 | 15±8 | 15±8 | 15±8 | |
| | Horizontal | 10 | 8±5 | 7±4 | 8±4 | 8±4 | 8±4 | 8±4 | |
| 50 (21 vol.%) | Up | 10 | 51±35 | 51±33 | 41±26 | 34±19 | 40±21 | 33±16 | 42±9 |
| | Down | 10 | 16±12 | 16±10 | 16±9 | 18±8 | 18±9 | 18±8 | |
| | Horizontal | 10 | 16±9 | 15±8 | 14±8 | 11±7 | 11±6 | 12±7 | |

## 3.2. Microstructure

### 3.2.1. Rayleigh-Bénard convective cells

**Fig. 2** shows polished cross-sections of sintered TiO$_2$ specimens obtained by solidifying suspensions with 40 wt.% TiO$_2$: (a) upward, (b) downward, and (c) horizontally, with respect to gravity. The center of each image corresponds roughly to the center of each respective sintered sample, *i.e.*, the images extend over the full diameter and height of the sintered samples. Overall dimensions of the samples shown in **Fig. 2** are: (a) 9 mm diameter, 2.5 mm high; (b) 10 mm diameter; 2.2 mm high; (c) 9.6 mm diameter; 2.1 mm high. The reduction in diameter/height from the original solidified dimensions (15 mm diameter; 3 mm height) are attributed to sintering shrinkage and error. Error related to cutting and polishing samples to obtain images that correspond exactly to the center of samples accounts for a reduction of ~2 mm of the true sample diameter (*i.e.*, the center of images shown are located within ~2 mm of the true geometric center of each sample).

The images in **Fig. 2**(b) and (c) are rotated such that the first- and last-to-solidify regions of each image are shown at the bottom and top, respectively, for all three cross-sections, **Fig. 2** (a-c). Significant misalignment of the wall orientation (*i.e*., the ice dendrite orientation) with respect to the imposed temperature gradient is observed for the sample solidified upward, **Fig. 2**(a). Considering a maximum angle of 90° as "full alignment" (wall orientation is parallel to the induced thermal gradient), tilt angles ($\theta$) as low as 50 to 65° with respect to the base of the sample are observed. Misalignment directions correspond to two semicircular regions (red dotted arrows in **Fig. 2**-a) with approximate mirror symmetry about the center line (vertical red dotted line in **Fig. 2**-a). This pattern was consistently observed in nearly all 20 cross-sections of 30 and 40 wt.% $TiO_2$ suspensions solidified upward (the absence of this pattern for 50 wt.% $TiO_2$ suspensions solidified upward is attributed to ice lensing, which is discussed in the following). Minimal misalignment (~70 to 89°) was observed for downward (**Fig. 2**-b) and horizontal (**Fig. 2**-c) solidification. Moreover, no consistent pattern of misalignment was observed across specimens solidified downward nor horizontally (**Fig. 2**-b and c). The location and direction of misalignment, when it existed (~50% of samples), appeared to be random, and may have resulted from post-solidification processing steps (*i.e*., anisotropic sintering shrinkage).

Tankin *et al*. [50] investigated the upward and downward solidification of water using a Mach-Zehnder interferometer. For the upward solidification case, the cold plate temperature was reduced from 23 to -5°C, while the top of the fluid was held at ~23°C (the solidification velocity was not reported). Immediately upon cooling, a parallel row of Rayleigh-Bénard convective cells was observed in the fluid region closest to the cold source; individual cells extended vertically upward from the cold source to the ~6°C isotherm (2°C higher than the density maximum of water). With continued cooling and initiation of freezing, the size of individual cells increased while the total number of cells decreased (from three to two cells). An interferogram [50] showing the parallel row of convective cells observed immediately after onset of freezing is shown in **Fig. 3**(a). The red, dashed line in (a) indicates the approximate position of the solid/liquid interface (which advanced upward in the direction of the induced temperature gradient, *T*); the convective cells are located immediately above the solid/liquid interface. As shown in **Fig. 3**(b), convective cells were observed for the downward solidification configuration as well. Here, the solid/liquid interface is located at the top of the image and solidification proceeds downward. For downward solidification, the row of convective cells was confined to the region where the fluid temperature was ~4°C or greater; thus, they did not appear to interact directly with the solidification front. A similar observation was made by Brewster *et al.* [51], for the downward solidification of water that contained tracer particles (40 $\mu$m in diameter).

The microstructural pattern present in **Fig. 2**(a) for upward solidification of aqueous suspensions of 40 wt.% $TiO_2$ is consistent with the convection pattern observed by Tankin *et al*. [50], for the upward solidification of water (**Fig. 3**-a). Assuming that the microstructural pattern shown in **Fig. 2**(a) is an imprint of solidification events, it can be concluded that at least two Rayleigh-Bénard convective cells were present during solidification of our samples. However, since these structures are visualized after solidification, the microstructural imprint is not reflective of the convective cells themselves, but rather a consequence of their interaction with the solid/liquid interface. Further, it is possible that post-solidification processing steps (*i.e*., sublimation and sintering) may have introduced microstructural changes to the structures templated during solidification. Lastly, additional cells may have been identifiable with three-dimensional imaging techniques. *In-situ* observations of the solidification process, *e.g.*, *via* synchrotron x-ray tomography, may help clarify these observations.

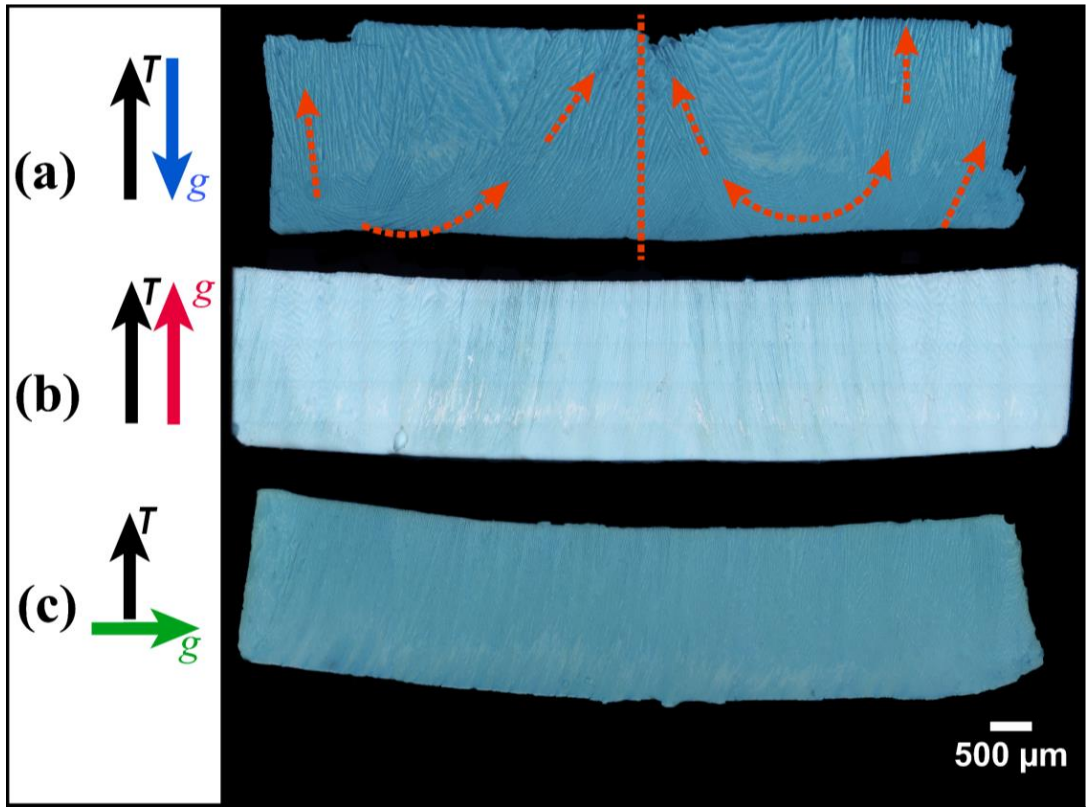

**Fig. 2.** Optical micrographs of polished longitudinal cross-sections of sintered, freeze-cast $TiO_2$ obtained by directional solidification of aqueous suspensions containing 40 wt.% $TiO_2$ nanoparticles, where solidification direction is: (a) upward, (b) downward, and (c) horizontal with respect to the gravity vector. Images in (b) and (c) are rotated such that the first- and last-to-solidify regions of the sample are shown at the bottom and top of all images. Here, $T$ and $g$ are the directions of the temperature gradient and gravity, respectively, during solidification. The full height (3 mm, minus sintering shrinkage of ~25%) and full diameter (±2 mm) of each sample are shown.

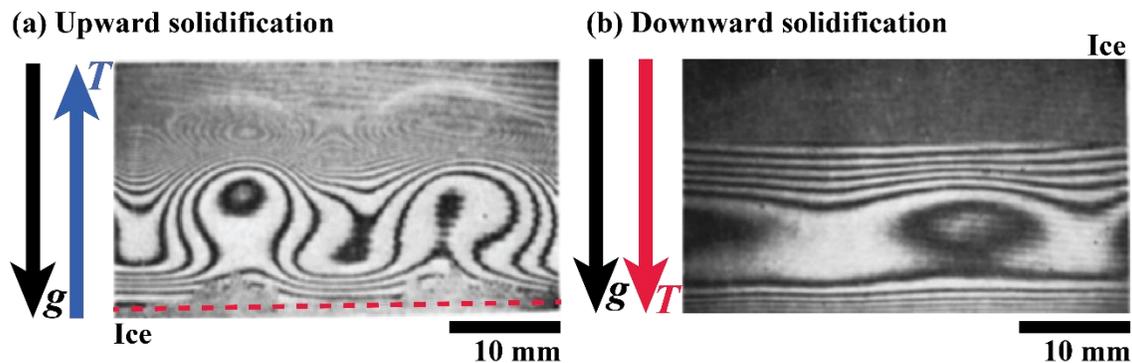

**Fig. 3.** Interferograms from Ref. [50], showing convective currents during (a) upward and (b) downward solidification of water, where $v$ indicates the direction of freezing and the red, dashed line in (a) marks the approximate position of the solid/liquid interface. In (b), the solid/liquid interface is located near the top of the image.

### 3.2.2. Interdendritic fluid flow

As discussed above, significant lamellar misalignment was observed for samples solidified upward, but not for samples solidified downward or horizontally. Deville *et al.* [27, 52, 53] suggest that microstructural misalignment in ice-templated materials results from an imbalance between the anisotropic growth kinetics of hexagonal ice and the induced thermal gradient. Growth along the basal face of ice (typically oriented perpendicular to the direction of heat flow [54-56]) is about 31 and 43% slower than growth along the prism and secondary prism faces, respectively [57]. This mechanism should be independent of the direction of solidification. Given that we did not observe significant tilting in neither downward nor horizontal solidification configurations, we propose an alternative explanation for the tilting observed here (for the upward solidification of aqueous $TiO_2$ at the high solidification velocities, 75±10 μm·s$^{-1}$).

The tilting shown in **Fig. 2**(a) is consistent with the so-called "upstream effect" described in seawater [58-61] and alloy solidification [62-64] literature, wherein dendrites tilt in the direction of buoyancy-driven convective fluid motion [59] arising from density gradients in temperature and/or composition [65]. The upward solidification of aqueous $TiO_2$ suspensions can be directly compared to the downward solidification of sea ice because, in both cases, the liquid at the solid/liquid interface is buoyant (due to the salinity of seawater, a density inversion in the liquid is not observed during solidification at typical undercoolings [66, 67]). Thus, the liquid at the solid/liquid interface is denser than the bulk liquid below, promoting buoyancy-driven, interdendritic fluid motion (the warmer, less dense seawater from below flows upward and displaces the colder, denser seawater at the interface). Phase-field modeling and subsequent experimental work on the upward solidification of a succinonitrile-3.5 wt.% $H_2O$ alloy melt with forced, interdendritic flow (induced via shear flow initiated by a rotating magnetic field) revealed a similar effect [68]. The upstream effect results from interdendritic fluid motion, *i.e.,* fluid motion that is largely confined to the mushy region (the region between the fully solidified solid and the fully liquid suspension, where solid ice and liquid suspension coexist [69]). Interdendritic fluid motion can be driven by the shear flow (across the solid/liquid interface) produced by macroscopic convective fluid motion in the bulk liquid, ahead of the solid/liquid interface (described here as a row of two Rayleigh-Bénard convective cells). It can also occur independently of the macroscopic convective pattern, with buoyancy remaining the driving force [69]. In either case, interdendritic fluid flow is weaker than flow resulting from the macroscopic convection described above (approximately one-thousandth that of the bulk fluid region [70]).

Interdendritic fluid flow also explains observations of asymmetric, one-sided dendritic arms commonly reported in ice-templating studies for both particle suspensions [3, 18, 27, 49, 52, 53, 71-73] and polymer solutions [74, 75]: secondary arm growth is suppressed on the "downstream" side [68, 76, 77]. This is depicted in **Fig. 4** where (a) shows the "ideal" case, where no appreciable interdendritic flow exists and (b) shows the effect of buoyancy-driven, interdendritic flow for upward solidification. Here, shear flow is depicted to be produced due to the presence of macroscopic convective cells located within the bulk suspension region (ahead of the solid/liquid interface) and the direction of the shear flow is consistent with a concave interface curvature (justified in the following). Consequences of interdendritic flow include: (i) tilting of ice lamellae and (ii) suppression of secondary dendrite arms on the "downstream" side. The flow direction promotes growth of dendritic arms on the upstream side, where heat is more easily transported away from the lamellae (*via* fluid flow) than on the downstream side. The warmer fluid on the downstream side promotes melting and fragmentation of secondary dendritic arms [64].

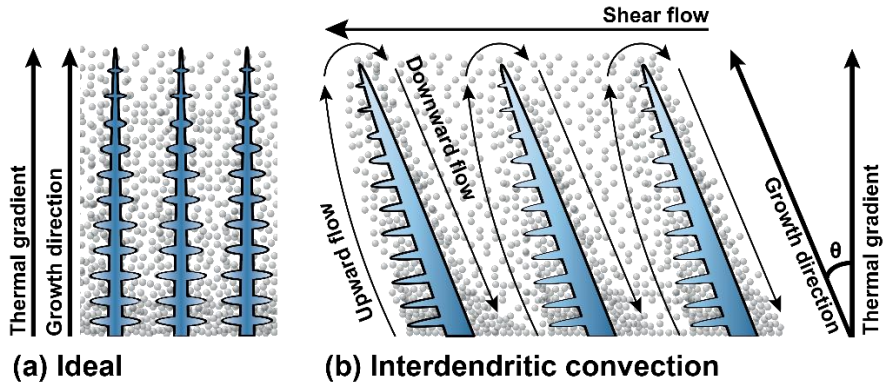

**Fig. 4.** Schematic showing the effect of interdendritic fluid flow on morphology of ice (blue) dendrites, where (a) no appreciable interdendritic fluid motion is present during solidification, and (b) there is buoyancy-driven convection, resulting in: (i) tilting of ice lamellae and (ii) suppression of secondary dendritic arms on the "downstream" side of ice lamellae.

**Fig. 5** shows SEM images of polished cross-sections of sintered $TiO_2$ obtained after upward solidification of a 30 wt.% suspension. The overall microstructural pattern is similar to that shown in **Fig. 2**(a) for 40 wt.% $TiO_2$ solidified upward; *i.e.*, two semicircular regions with approximate mirror symmetry about the center line were observed (an optical micrograph depicting the entire cross-section is provided as supplementary material, Fig. S-3). In terms of **Fig. 2**(a), the image shown in **Fig. 5**(a) is obtained from the left-hand side of the image, whereas (b) and (c) correspond to the central-right and right-hand side, respectively. **Fig. 5**(d) and (e) are lower magnification images of (a) and (c), respectively. In all images, epoxy-filled pores are visible as dark regions and sintered, $TiO_2$ particle walls as light regions. Evidence of asymmetric, secondary dendritic arms is observed on the right-hand side of particle walls in **Figs. 5**(a,d) and on the left-hand side in **Figs. 5**(c, e; some are marked with yellow arrows), which is consistent with the upstream effect depicted in **Fig. 4**. In **Fig. 5**(a, d), secondary dendritic arms are observed on the right-hand side of TiO$_2$ walls, which corresponds to the left-hand side of ice dendrites (represented by epoxy, shown as dark regions); thus, secondary dendritic arms are suppressed on the right-hand, downstream side of ice dendrites, as shown in the schematic below **Fig. 5**(a). In **Fig. 5**(c, e), the direction of dendrite tilt changes, and the schematic below **Fig. 5**(c) shows the corresponding change in fluid flow direction. Secondary dendritic arms are now absent on the left-hand side of ice dendrites, which still corresponds to the downstream side; secondary arms are present on the right-hand, upstream side.

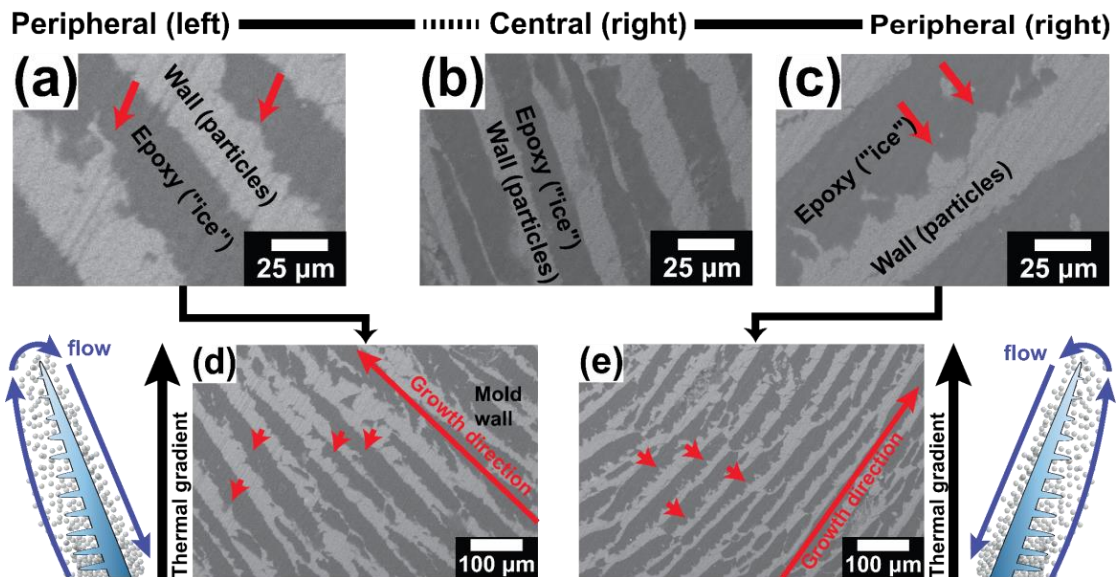

**Fig. 5.** SEM micrographs of polished longitudinal cross-sections of sintered, freeze-cast TiO$_2$ obtained by solidifying aqueous suspensions containing 30 wt.% TiO$_2$ nanoparticles vertically upward. Images in the top row are obtained from the: (a) left-peripheral (near mold wall), (b) central-right, and (c) right-peripheral (near mold wall) regions of the sample; (d) and (e) are lower magnification images of (a) and (c) respectively. Epoxy-filled pores are the dark regions of the samples and sintered TiO$_2$ particle walls are the light regions. Asymmetric secondary dendritic arms (yellow arrows) are observed in (a, d) and (c, e) where the ice growth direction diverges significantly from the direction of the imposed thermal gradient; these features are minimized in (b) where the tilt angle ($\theta$; ice growth direction) is minimized.

A significant difference in spacing ($\lambda_2$) between secondary dendrite arms was not found when comparing measurements between the left and the right regions (28±15 $\mu$m and 28±16, respectively). Similarly, mean secondary arm length did not vary between left and right regions (11±4 $\mu$m for both; **Table 2**). Relative to the images shown in (a) and (c), asymmetric dendritic arms and microstructural tilting are suppressed in (b), which corresponds to the central region of the sample. This is consistent with a reduction of interdendritic flow in the center portion of samples, where dendrites are better aligned with the induced thermal gradient, relative to the peripheral regions where dendrites are tilted with respect to the induced thermal gradient.

**Table 2.** Summary of primary and secondary dendrite spacing and length, based on SEM investigation of 9 samples (N = number of samples) created from aqueous suspensions with 30% wt.% TiO$_2$ solidified against ("Up"), parallel to ("Down"), and perpendicular to ("Horizontal") the gravity vector.

| Direction | $N$ | Primary spacing ($\lambda_1$, μm) | Secondary spacing ($\lambda_2$, μm) | | | Secondary arm length (μm) | | |
|---|---|---|---|---|---|---|---|---|
| | | | Overall | Left | Right | Overall | Left | Right |
| Up | 3 | 25±17 | 28±15 | 28±16 | - | 11±4 | 11±4 | - |
| | | | | - | 25±11 | | - | 10±5 |
| Down | 3 | 14±7 | 8±4 | 8±4 | 8±3 | 3±1 | 3±1 | 3±1 |
| Horizontal | 3 | 8±2 | 10±4 | 10±3 | 10±4 | 3±1 | 3±1 | 3±1 |

**Fig. 6** shows SEM images of polished cross-sections of sintered TiO$_2$ obtained by solidifying suspensions with a 30 wt.% initial particle fraction (a, c) downward and (b, d) horizontally. Short, secondary dendritic arms are observed on both sides of the sintered-particle walls (yellow arrows). Statistically-significant differences in terms of secondary arm spacing or length were not found when comparing those measured on the right-hand side to those measured on the left-hand side of dendritic walls for downward nor horizontal solidification (**Table 2**).

Bridging between particle walls is observed throughout these samples; some bridges are marked by yellow arrows in **Fig. 6**(c,d). These bridges appear similar to those observed by Cheng *et al.* [78, 79] for the directional freeze-casting of hydroxyapatite particles suspended in water containing $H_2O_2$ (where $H_2O_2$ introduces gas bubbles in the suspension and microporosity in sintered samples); thus, these bridges may have occurred as a result of residual gas within the suspension. As described in terms of **Fig. S-2**, whereas air bubbles rise to the suspension surface during upward solidification, they collect at the solid/liquid interface during downward and horizontal solidification.

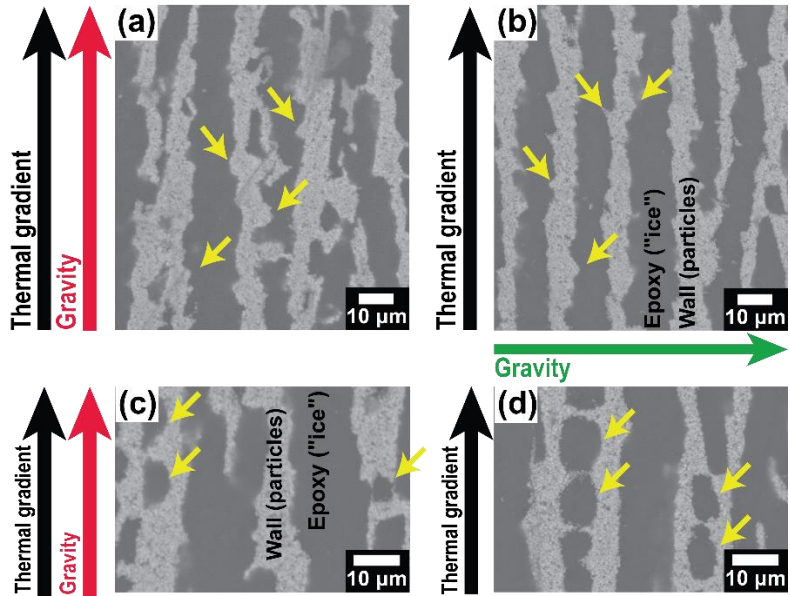

**Fig. 6**. SEM micrographs of polished longitudinal cross-sections of sintered, freeze-cast $TiO_2$ obtained by solidifying aqueous suspensions containing 30 wt.% $TiO_2$ nanoparticles: (a, c) vertically downward and (b, d) horizontally with respect to gravity. Images in (b) and (c) are rotated such that the "base" and "tops" of the samples (first- and last-to-solidify, respectively) are shown at the bottom and top of all images. Epoxy-filled pores are the dark regions of the samples whereas sintered $TiO_2$ walls are the light regions. Minimal microstructural tilting (divergence between the imposed thermal gradient and the dendrite growth direction) is observed and secondary dendritic arms are relatively symmetric (marked by yellow arrows). Micrographs in (c) and (d) show microstructural bridging (marked by yellow arrows) in both downward and horizontally solidified samples.

Our results are consistent with a reduction (or suppression) of fluid flow parallel to the solidification interface (*i.e.*, shear flow in **Fig. 4**-b) for downward and horizontal solidification. In these cases, symmetric, secondary arms are present on both sides of lamellae and primary arms are not tilted with respect to the induced thermal gradient. By contrast, tilted lamellae and asymmetric secondary arms are observed for upward solidification.

### *3.2.3.   Ice lens defects*

Ice lenses are the most commonly observed defects in ice-templated materials and significantly compromise their mechanical properties [32]. Ice lenses arise during solidification as ice-filled platelets, oriented perpendicular to the freezing direction, and are templated into cracks after sublimation. Ice lensing is studied in a variety of fields, *e.g.,* geology (frost heaves [80]), food engineering [81], and cryobiology [82]. Although various hypotheses have been proposed to explain their development [41, 83-88], most of the ice-templating literature attributes ice lens formation to particle engulfment, rather than particle rejection, at the solidification interface [2, 37-40]. During directional solidification, particles are first rejected by the advancing solid/liquid interface before they are incorporated within interdendritic space. As a result, a region of accumulated particles forms in the bulk suspension, directly ahead of the solid/liquid interface. If the particle fraction within this accumulation region is far enough below the breakthrough concentration (described above) [42], ice lamellae or dendritic structures form (**Fig. 7**-a,b). Above the breakthrough concentration, lamellae/dendrites are unable to propagate, and ice lenses may develop. Two distinct ice lensing regimes can be identified from the literature; these are described by You *et al.* [41] as ice spears (**Fig. 7**-c) and ice bands **Fig. 7**-d).

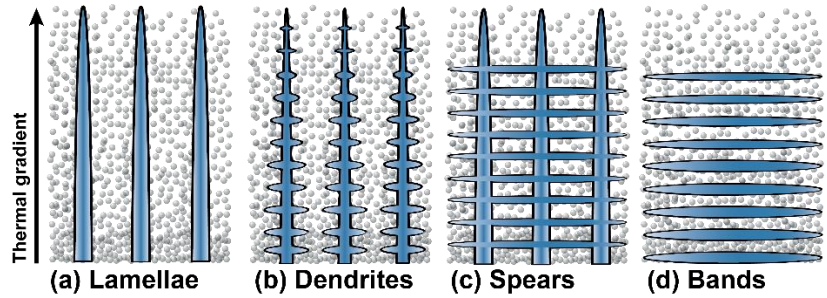

**Fig. 7.** Schematic showing the morphology transition of ice between (a) lamellar and/or (b) dendritic structures to ice lensing regimes, including (c) ice spears and (d) ice bands.

Ice bands develop when a fraction of particles within the accumulation region is engulfed by the solid/liquid interface (rather than incorporated within interdendritic space), while other particles are pushed ahead of the front [84]. The engulfed particles form a discrete, enriched particle layer (parallel to the solidification front), and a particle-free ice layer (*i.e.*, "ice band") forms immediately thereafter (*e.g.*, Fig. 1-a of Ref. [89] and Fig. 5 of Ref. [43]). The particle accumulation region is still present after the first ice lens develops and contains particles that were initially pushed during the development of the ice lens as well as additional particles that entered the region through other mechanisms (*e.g.*, sedimentation). The periodic ice banding process may repeat indefinitely, provided water and particles are available for incorporation. The main distinction between the ice banding and ice spear regimes (**Fig. 8**-c), is that the former tends to form discrete bands of ice (and enriched particle layers) whereas the latter does not. For periodic banding (**Fig. 8**-d), ice dendrites and particle-packed walls that are oriented parallel to the freezing direction are absent. In the ice-spear regime (**Fig. 8**-c), alternating regions of pores and walls oriented parallel to the freezing direction are observed. However, in sintered samples, the walls contain microcracks that are oriented perpendicular to the freezing direction (*e.g.*, Fig. 9-b of Ref. [33] and Fig. 2-b of ref. [37]). We believe that a possible explanation for the ice-spear regime is that, near the breakthrough concentration, water is forced into the already formed particle-packed walls, creating pure-ice-filled layers (oriented perpendicular to the freezing direction) within the vertical wall structures, while the vertical, lamellar ice region remains intact. Peppin *et al.* [39] observed a transition between these regimes where periodic banding ice-lenses were observed for aqueous suspensions containing 74 wt.% (~52 vol.%) kaolinite clay, while a decrease in the initial particle fraction to 50 wt.% (~27 vol.%) produced a structure consistent with the ice-spear regime.

Ice lens defects were observed in all samples solidified upward, for all particle fractions under study ($N = 30$). Similar defects were not observed in any samples solidified downward ($N = 30$) nor horizontally ($N = 30$). Given that thermal buoyancy flow is stabilized for downward solidification and partially stabilized for horizontal solidification, these observations provide strong evidence that buoyancy-driven fluid flow increases the propensity of ice lens formation during upward solidification (as discussed in the introduction, Stoke's sedimentation velocity of nanometric spheres in water is negligible). These observations are also consistent with results which we reported previously [43], wherein ice lens defects were observed for 20 wt.% $TiO_2$ suspensions solidified upward under normal terrestrial gravity ($1g$) but were not observed for suspensions solidified upward under reduced gravity (where buoyancy-driven fluid motion is reduced).

**Fig. 8** shows representative images of ice lens defects in sintered $TiO_2$ for initial particle fractions of: (a) 30, (b) 40, and (c) 50 wt.%. The defects are visible as black, horizontal lines (cracks, which were previously ice) that stretch across the white, $TiO_2$ walls and are easier to discern in the higher magnification images provided in the second row of **Fig. 8**. Most cases show horizontal micro-cracks which nearly span the width of individual sintered $TiO_2$ walls; these micro-cracks are consistent with the ice-spear regime discussed above. Crack width (and wall width) is observed to increase with increasing: (i) initial particle fraction in the suspension, (ii) vertical distance from the first-to-solidify region (in contact with the aluminum mold base) to the last-to-solidify region, and (iii) radial distance from the outer wall to the center of the specimen. Dividing samples in half horizontally, mean crack width for 30 wt.% $TiO_2$, is 30±12 and 56±24 μm in the first- and last-to-solidify regions, respectively. At 40 and 50 wt.% $TiO_2$, these values increase to 32±14 and 67±23 μm, respectively, within the first-to-solidify regions and to 66±15 and 197±95 μm for the last-to-solidify regions (for 40 and 50 wt.% $TiO_2$, respectively). Additionally, (iv) the vertical distance between the base of the sample (in contact with the freezing aluminum mold base) and the first observation of an ice lens defect decreases from 267±40 μm for 30 wt.% $TiO_2$ to 202±24 and 69±14 μm for 40 and 50 wt.% $TiO_2$, respectively (see supplementary Fig. S-4). That is, ice lens defects develop earlier in the solidification process for suspensions containing a higher initial particle fraction. We show in the following that factors (i) through (iv) lend support to the argument that fluid motion plays a role in initiating a pattern of particle engulfment at the solid/liquid interface that leads to the formation of ice lens defects. This is most apparent when considering the 50 wt.% $TiO_2$ samples where ice lens defects are pervasive.

**Fig. 9** shows cross-sectional images of sintered $TiO_2$ solidified upward using the highest particle fraction of 50 wt.% $TiO_2$. **Fig. 9**(a) is taken from the center and (b) is obtained from the outermost portion of the sample (the right side of the image corresponds to the peripheral region of the sample in contact with the solidification mold; total length of the image is ~4.5 mm). Ice lens defects are visible as black, horizontal lines (cracks) that stretch across the white, $TiO_2$ walls.

As noted above, crack (and wall) width increases from the base (first-to-solidify region) to the top of the samples; increases of 87 and 194% were noted for 30 and 50 wt.% $TiO_2$, respectively. The higher magnification images (offset in red boxes) convey that the disparity between 30 (**Fig. 8**-a) and 50 wt.% $TiO_2$ occurs because of widespread merging of walls (where two or more walls merge into one) in the higher particle fraction samples. This may represent a transition from the ice spear to the periodic banding regime of ice lensing discussed above.

Wall merging occurs at lower sample heights within the central region of 50 wt.% $TiO_2$ samples (**Fig. 9**-a) than in the outer regions (**Fig. 9**-b). Considering merging of walls that occurs at heights greater than 150 μm from the base of 50 wt.% $TiO_2$ samples, the mean height at which wall merging is observed is 254±79 and 463±135 μm (from the sample base) for central and peripheral regions, respectively. Within the uppermost region of the 50 wt.% $TiO_2$ sample depicted in **Fig. 9**, a polygonal pattern develops. A similar pattern was observed by Peppin *et al.* [89, 90] for the upward solidification of 50 wt.% kaolinite powders (~28 vol.%) in water. Green bodies are particularly fragile within this region and portions of samples break off during post-solidification processing. This can be inferred by the jagged top surface of the sample shown in. The shorter overall height of the central (a) *vs.* peripheral (b) images is likely a consequence of this pattern beginning earlier during solidification within the central region of the sample than within peripheral regions (*i.e.*, sample losses during post-solidification processing are greatest in the central region of samples).

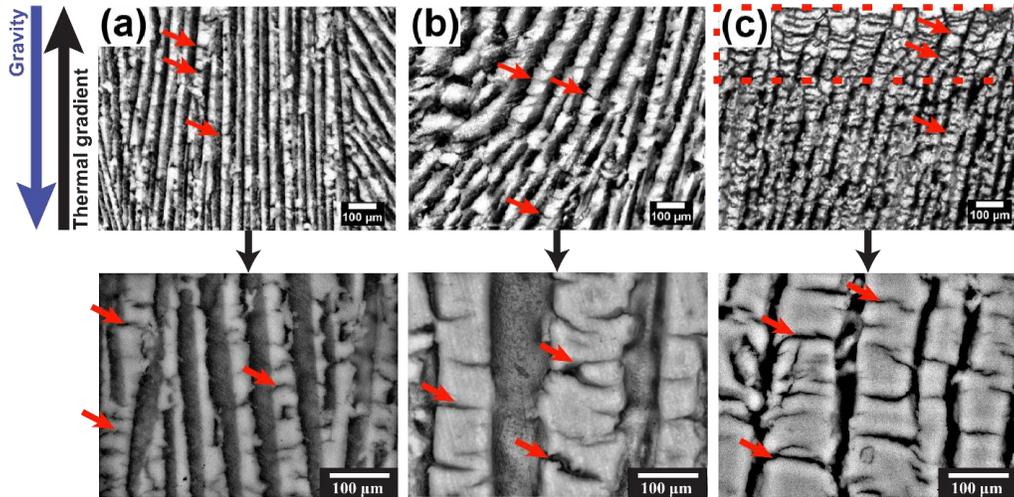

**Fig. 8.** Optical micrographs of polished longitudinal cross-sections of sintered, freeze-cast $TiO_2$ obtained by directional solidification of aqueous suspensions containing (a) 30, (b) 40, and (c) 50 wt.% $TiO_2$ nano-particles, where solidification direction is upward (against gravity) for all. Higher magnification micrographs are shown in the second row and correspond to the images above as indicated by black arrows; *v* and *g* are the directions of velocity and gravity, respectively, during solidification. Pores and walls are shown in black and white, respectively. Two regimes of ice lens defects are observed. Ice spear defects are visible as thin black microcracks (red arrows, a-c) oriented perpendicular to the freezing direction, stretching across the white $TiO_2$ walls. The onset of ice band defects is observed toward the top of the image in (c; red box); here, wall merging is observed, and crack length increases in the wider walls.

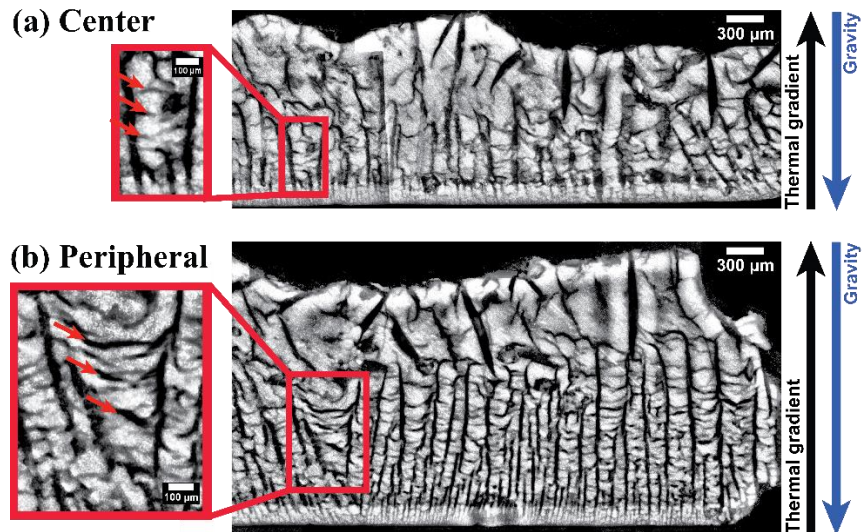

**Fig. 9.** Optical micrographs of polished longitudinal cross-sections of sintered, freeze-cast $TiO_2$ showing pervasive ice lens defects in a sample solidified upward using the highest initial particle fraction of 50 wt.% $TiO_2$ particles. The top image (a) is taken from the center of the sample, whereas the bottom image (b) is obtained from the outermost portion of the sample (contact surface with PVC mold is shown on the right in (b)). Arrows indicate directions of (*v*) solidification and (*g*) gravity. Pores and walls are shown in black and white, respectively. Ice lens defects are visible as thin black cracks (a few are highlighted with red arrows) oriented perpendicular to the freezing direction, stretching across the white $TiO_2$ sintered walls.

**Fig. 10** shows optical micrographs of polished cross-sections of sintered materials obtained by directionally solidifying 30, 40, and 50 wt.% TiO$_2$ suspensions (columns 1-2, 3-4, and 5-6, respectively), where solidification direction is: (blue) upward, (red) downward, and (green) horizontally with respect to the gravity vector (g). Images in (b) and (c) are rotated such that the "base" and "tops" of the samples (first- and last-to-solidify, respectively) are shown at the bottom and top of all images. Images shown in the first and second columns of each particle fraction were obtained from the outer "peripheral" and central region of each sample (the horizontal line observed in 50 wt.% TiO$_2$ upward and horizontal solidification images is an artifact resulting from image stitching). **Fig. 10** shows that fine, lamellar microstructures can be obtained for up to 50 wt.% TiO$_2$ when samples are solidified downward or horizontally, but not when they are solidified vertically upward.

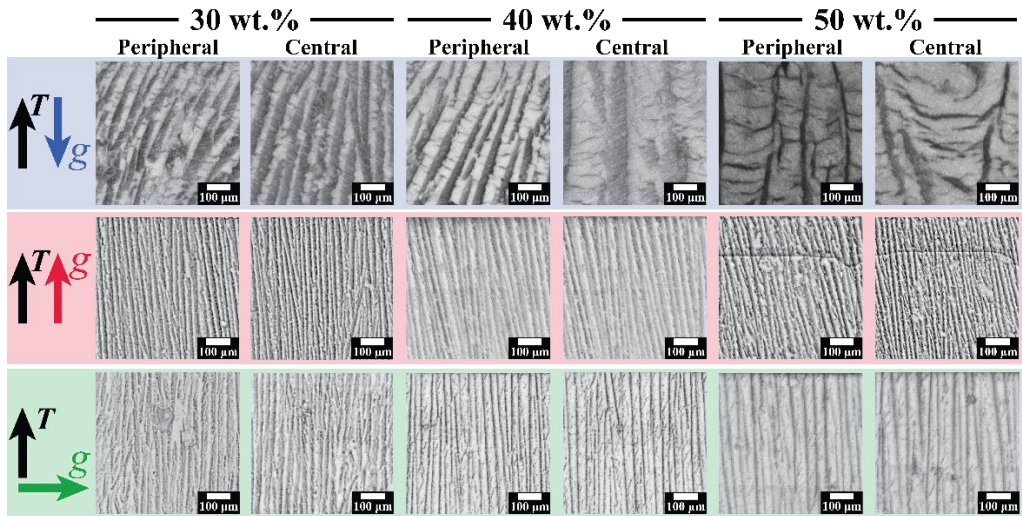

**Fig. 10.** Optical micrographs of polished longitudinal cross-sections of sintered, freeze-cast TiO$_2$ obtained by directionally solidifying suspensions with 30, 40, and 50 wt.% TiO$_2$ nanoparticles (columns 1-2, 3-4, and 5-6, respectively), where solidification ($v$) is: (blue) upward, (red) downward, and (green) horizontally with respect to the gravity vector (g). Images in (b) and (c) are rotated such that the "base" and "tops" of the samples (first- and last-to-solidify, respectively) are shown at the bottom and top of all images. Images shown in the first and second columns of each particle fraction were obtained from the outer "peripheral" and central regions of each sample, respectively. The horizontal line visible in 50 wt.% in images for upward and horizontal solidification is an artifact resulting from image stitching.

### 3.2.4. Radial segregation

**Fig. 11** shows mean pore width and wall width as a function of TiO$_2$ weight fraction (30, 40, and 50%) for samples solidified (a, d) upward (blue), (b, e) downward (red), and (c, f) horizontally (green). In comparison to downward and horizontal solidification, an increase in pore width and wall width is observed in samples solidified upward for all TiO$_2$ weight fractions. This effect is greatest for 30 wt.% TiO$_2$, where mean pore width is about four times larger when solidified upward compared to downward and about three times as large in comparison to horizontal solidification (27±22, 7±5, and 10±6 μm for the up, down, and horizontal solidification directions, respectively; **Table 1**). At 50 wt.% TiO$_2$, mean pore width is approximately three times as large for upward solidification as compared to horizontal and downward solidification (51±35, 16±12, and 16±9 μm for the up, down, and horizontal solidification, respectively; **Table 1**). The same ratios are likewise observed for wall width. We attribute these disparities to the formation of ice lenses (which cause individual walls to merge into larger walls), as discussed previously.

In a previous study [43], we observed a macroscopically concave solid/liquid interface curvature, with a depression in the center of the sample, during upward solidification of aqueous suspensions of 5-20 wt.% TiO$_2$; moreover, the depth of the depression was observed to increase as solidification progressed. The freezing set-up utilized here did not allow for observation of the interface during solidification. Nevertheless, the macroscopic microstructural pattern observed for upward solidification (**Fig. 2**-a, two semicircular regions with approximate mirror symmetry about the center line) is consistent with the macroscopic convective fluid flow pattern we described previously [43]. Interface curvature can result from, among other factors, a mismatch between thermal conductivities of liquid, solid, and mold. For example, if mold walls are more thermally conductive than solid ice, the interface is expected to be concave because latent heat of solidification is preferentially evacuated through the mold wall [62]. Here (and previously [43]), the thermal conductivity of the PVC mold (~0.2 W/m K) is much lower than that of the ice (~2.2 W/m K); thus, if a thermal conductivity mismatch was responsible for interface curvature observed previously (and inferred here), we would expect the curvature to be convex, rather than concave. We posit that the concave curvature observed previously (and inferred here) results from the macroscopic convection pattern itself.

A disparity in mean pore width between the outer edge of the sample and its center is expected when a curvature of the solid/liquid interface is present. This so called "radial segregation" occurs because solidification velocity is inversely proportional to the thickness of ice lamellae; faster solidification velocities tend to result in thinner pores [52], and a concave curvature implies solidification velocity is faster at the sides of the mold than the center [91]. In **Fig. 11**, dark markers represent mean values of pore and wall width taken from the center of samples; light markers denote mean peripheral measurements (as discussed above, longitudinal cross-sections are divided in three horizontally; peripheral measurements refer to those obtained from the two outside images and central measurements are taken from the center image, see Fig. S1). For upward solidification, widths of both pores and walls are larger in the center of samples than near the mold, for all $TiO_2$ weight fractions. This effect is highest at 30 wt.% $TiO_2$, where mean pore width measured from central and peripheral regions are 39±27 and 19±12 μm, respectively (corresponding to mean wall width from central and peripheral regions of 56±37 and 30±22 μm, respectively). In contrast, for upward and downward solidification, mean values of pore and wall width are nearly the same between central and peripheral locations (statistically significant differences between these measurements were not found). In other words, we find evidence of radial segregation in the case of upward solidification, but not in downward or horizontal configurations.

In the alloy literature, axial rotation [91] and/or vibration [92] is sometimes used to counterbalance buoyancy-driven fluid flow which would otherwise lead to radial segregation. For example, for the upward solidification of succinonitrile-0.005 wt.% ethanol alloys, an axial rotation rate of 150 rpm was used to prevent radial segregation [91]. However, for upward solidification of hypereutectic Sn-25Pb where Sn, the species rejected in the melt at the solid/liquid interface, is less dense than the melt and thus buoyant), Grugel *et al*. [93] observed radial segregation regardless of any applied axial rotation; horizontal solidification with axial rotation was needed to offset buoyancy-driven convection and achieve uniformly aligned dendritic microstructures; downward solidification was not tested.

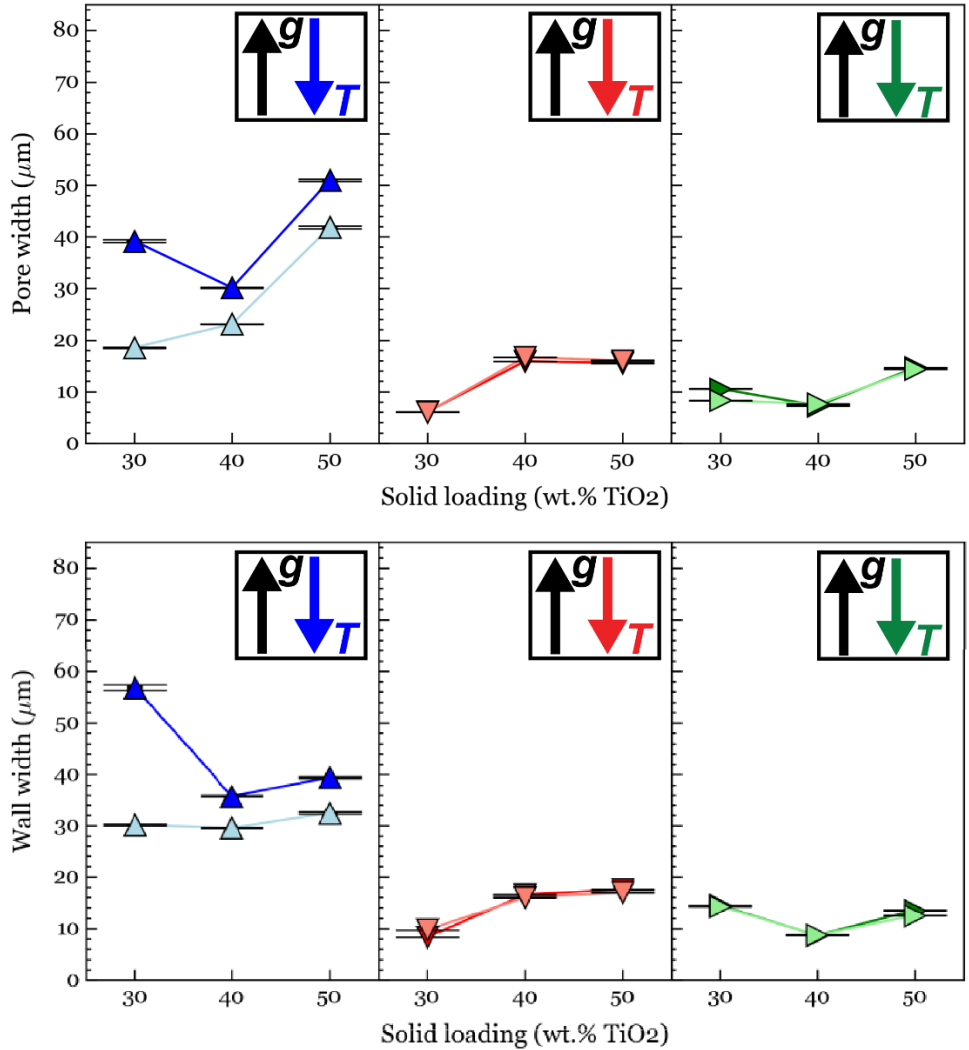

Fig. 11. Graphs showing widths of pores (top row) and walls (bottom row) in sintered, freeze-cast $TiO_2$ obtained by directional solidification of aqueous suspensions containing 30, 40, and 50 wt.% $TiO_2$ nanoparticles, where solidification direction is: (blue) upward, (red) downward, and (green) horizontal with respect to the gravity vector. The dark and light markers represent data collected in the "center" and "peripheral" regions of samples, respectively. Error bars are derived using a 99.95% confidence interval about the mean.

### 3.2.5. *Model comparison*

You *et al.* [41] developed a model for predicting freeze-cast morphologies based on solidification conditions and suspension characteristics. Morphologies predicted by the model include: dendritic, "ice spears", and "frozen fringe" (periodic banding). Fig. 12 shows a modified version of the phase diagram provided by these authors which includes data points calculated for our current experiments (star markers) and our previous work [43] (triangle markers). Predictions are based on the relationship between three dimensionless parameters identified by the authors; the Darcy coefficient (*D*), the film coefficient (*F*), and a dimensionless parameter Φ, which is given by:

$$\Phi = \frac{\phi_0}{\phi_p - \phi_0} \quad \text{Eq. 1}$$

where $\phi_0$ is the volume fraction of particles in the bulk suspension (here, $\phi_0$ =0.10, 0.15, and 0.21) and $\phi_p$ is the volume fraction of particles in the accumulation region, taken as the random close-packed value of 0.64. The dimensionless Darcy coefficient, *D*, describes fluid flow through the accumulated particle region and is expressed as:

$$D = \frac{\mu v T_m}{k \rho L G} \quad \text{Eq. 2}$$

where $\mu$ is the dynamic viscosity of the fluid (taken as 1.78 mPa·s for water at 0°C [94]), $v$ is solidification velocity (approximated here at 75 $\mu$m · s$^{-1}$), $T_M$ is the melting point of water, $\rho$ is the density of ice, $L$ is the latent heat of fusion of ice, $G$ is the temperature gradient (21 K· mm$^{-1}$), and $k$ is the permeability of the accumulated particle region, given by:

$$k = \frac{r^2(1-\phi_p)^3}{45\phi_p^2} \quad \text{Eq. 3}$$

where *r* is the particle radius (15 nm).

The dimensionless film coefficient, *F*, describes flow within the pre-melted liquid film between a particle and the solidification front and it is given by:

$$F = \frac{8\pi^2\sqrt{3}\mu v r^2}{A\lambda^3} \qquad \text{Eq. 4}$$

where *A* is the Hamaker constant between the solidified fluid and the particle (taken as $7 \cdot 10^{-20}$ J for $TiO_2$/ice [95]) and $\lambda$ is an empirical constant, given in Ref. [41] as 0.225.

Dendritic structures are predicted when $D/(1 + \Phi)$ is greater than unity, whereas ice lensing regimes are predicted when $D/(1 + \Phi)$ is less than unity. The star markers in **Fig. 12** represent values calculated based on the present work; specifically, $D/(1 + \Phi) = 1.7$, 1.9, and 2.1 for 10, 15, and 21 vol.% $TiO_2$ nanoparticles, respectively; *i.e.,* dendritic regimes are predicted for all volume fractions studied here. Dendritic structures were indeed observed for all samples solidified downward and horizontally, but spear and banding regimes were observed in our samples solidified vertically upward (for all weight fractions). Triangle markers represent values calculated from our previous work [43] for 5-20 wt.% $TiO_2$ nanoparticles, where $v = 100$ μm·$s^{-1}$. Again, dendritic structures are predicted, yet ice banding was observed for 20 wt.% samples that were solidified under normal terrestrial gravity (but not for samples solidified under reduced gravity). Square symbols from literature values are also shown in **Fig. 12**, including solid red and black markers [83] ($Al_2O_3$, $r$=160 nm, $G$=2.58 K·$cm^{-1}$, $\phi_0$= 0.27, $v$=0.5-10 μm·$s^{-1}$, $A$=3.67x$10^{-20}$ J) and red and blue slashed squares [41] ($Al_2O_3$, $r = 100$ nm, $G = 8.19$ K·$cm^{-1}$, $\phi_0$= 0.27, $v = 7.3$-34.6 μm·$s^{-1}$, $A = 3.67x10^{-20}$ J).

Transitions from dendritic to lensing regimes are predicted as the freezing velocity decreases, the thermal gradient increases, the particle volume fraction increases, or the particle radius increases [41]. This model does not account for systems with solute effects (*e.g.*, dissolved binder in the suspension). However, it is known that the particle fraction within the accumulation region (and by consequence, the propensity for ice lenses to develop) is influenced by suspension characteristics (*e.g.*, suspension stability [37, 96]; particle sedimentation) and solidification conditions (*e.g.*, solidification velocity [25, 83], and as we show here, solidification direction). Here, we posit that buoyancy-driven fluid motion is also a contributing factor for the development of ice lenses in samples solidified vertically upward.

In the case of a concave solid/liquid interface curvature (described above), the depth of the concave depression at the solidification front increases as more particles are collected within the depression. The formation of a concave interface depression is supported by the observation of ice lens defects being created closer to the base of 50 wt.% $TiO_2$ samples within central regions, as compared to peripheral regions; in other words, the "breakthrough" particle concentration is reached earlier in the solidification process within the central region of samples. This breakthrough concentration is typically taken at the close-packing fraction value of 0.64 [42, 89] and represents the point at which the osmotic pressure of the accumulated particle region exceeds the capillary pressure necessary to allow ice to invade pore space [42] where "pore space" refers here to the space between compacted particles within the liquid accumulation region.

The initial suspension particle fraction of 50 wt.% corresponds to 21 vol.% $TiO_2$, which is significantly lower than the breakthrough fraction of 64 vol.%. Nevertheless, ice lens defects were observed, on average, at a height of 69±14 μm from the base for samples containing initial particle fractions of 50 wt.% (21 vol.%). Considering particle rejection by the solidification front as the only mechanism for creating the particle accumulation region, it is impossible for the volume fraction of particles at the solid/liquid interface to triple over the short distance that the solidification advances ~70 μm (~2% of the suspension height). Most likely, (i) the particle fraction at the base of the suspension exceeded that of the bulk suspension prior to the onset of solidification (due to sedimentation) and/or (ii) particles were enriched to the solid/liquid interface by other means (in addition to rejection by the solid/liquid interface), as discussed below.

As discussed earlier, the experiments of Tankin *et al*. [50] showed that the onset of buoyancy-driven convective fluid motion occurs when a thermal gradient is imposed. Thus, fluid motion-induced particle redistribution likely begins before solidification commences. Although the sedimentation rate is negligible for nanometric particles (especially at the high solidification velocities explored here), the sedimentation rate for large aggregates of particles may not be. This issue was explored by Lasalle *et al*. [37] with respect to the use of unstable suspensions. In contrast, orthokinetic aggregation (*i.e.*, stirring-induced aggregation) occurs when particles are redistributed by fluid flow. In such case, the rate of interparticle collisions increases drastically over that which would be expected via Brownian mechanisms alone. As a result, orthokinetic aggregation produces aggregates rapidly and considerable variation in aggregate size is observed [97]. The production of aggregates of different sizes gives way to an additional mode of aggregation which results from differential sedimentation rates. That is, aggregates of different diameters settle at different rates. Consequently, the particle collision rate is increased further; *i.e.*, particles are provided with an increased number of aggregation opportunities. A combination of fluid motion-induced particle redistribution and orthokinetic aggregation (and consequent enhanced sedimentation) may have effectively increased the particle fraction at the base of the suspension prior to the commencement of solidification and may be a contributing factor for why the model developed by You *et al*. [41] accurately predicts the dendritic structures observed here for downward and horizontal solidification, but not the structures observed for upward solidification. If our hypotheses posed here are correct, the particle volume fraction at the interface would be higher than that predicted by the model (over the course of solidification); this would lead to increased values of D/(1+Φ) while $F$ remained unchanged; *i.e.*, in terms of **Fig. 12**, *y*-axis values would decrease, moving prediction values toward (or within) lensing regime boundaries.

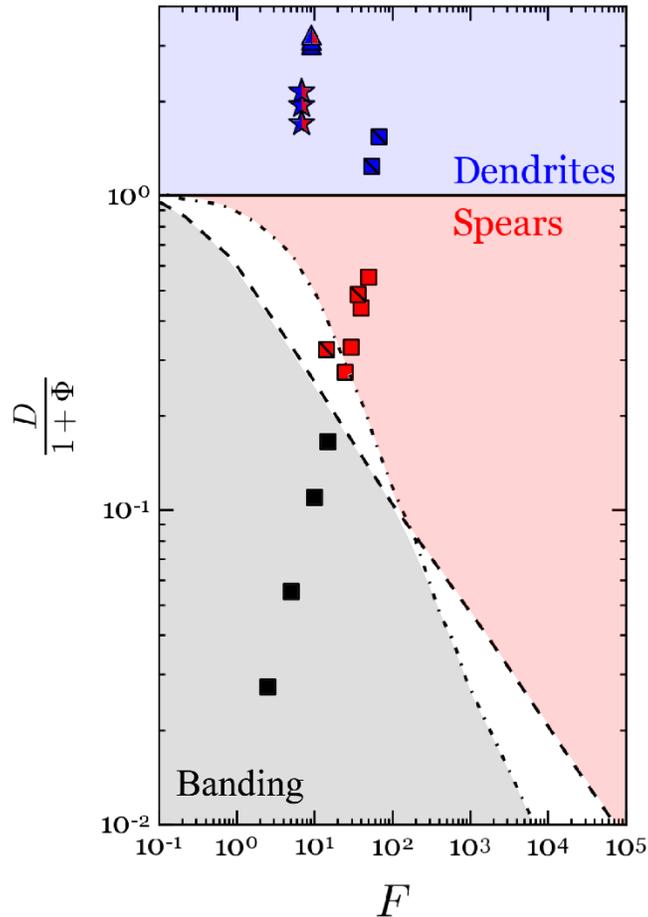

**Fig. 12.** Phase diagram after Ref. [41] showing predicted ice-templated morphologies, including: dendrites (blue), ice spears (red), and ice banding (grey) based on solidification and suspension parameters. The Darcy coefficient (*D*), film coefficient (*F*), and (Φ), are given by Eqs. 1-4. The star markers represent values calculated based on the present work; dendritic morphologies were observed for all TiO$_2$ weight fractions (30-50 wt.%; 10-21 vol.% TiO$_2$) for samples solidified downward and horizontally, whereas lensing regimes (ice lenses and spears) were observed in samples solidified upward (for all weight fractions). Triangle markers represent values measured in our previous work [43] for 5-20 wt.% (2-7 vol.%) TiO$_2$, where dendritic structures were observed for all weight fractions under microgravity and ice banding was observed for 20 wt.% TiO$_2$ solidified upward under normal terrestrial gravity. Square markers represent literature values, including solid red and black solid markers [83] (27 vol. % Al$_2$O$_3$, *r* = 160 nm ) and red and blue slashed markers [41] (27 vol.% Al$_2$O$_3$ *r* = 100 nm).

## 4. Conclusions

Ice-templated porous titanium dioxide ($TiO_2$) structures were fabricated by directional solidification (under thermal gradient) of aqueous suspensions containing 30, 40, and 50 wt.% $TiO_2$ nanoparticles in three different solidification configurations: (a) vertically upward (solid ice on bottom, liquid water on top), (b) vertically downward (liquid on bottom, solid on top), and (c) horizontal. Suspended particles were first rejected by the advancing solidification front and later incorporated within interdendritic space. After solidification, ice was removed via sublimation and the remaining particle-packed walls were sintered. Sintered structures consist of elongated, particle-packed walls separated by pore channels which template the ice dendrites.

Significant tilting of the wall orientation (*i.e.*, the ice dendrite misorientation) with respect to the temperature gradient was observed in samples solidified upward, but not for downward nor horizontal solidification. These results are consistent with: (i) the presence of a macroscopic convective fluid pattern during upward solidification and (ii) a reduction (or suppression) of interdendritic fluid flow in the cases of downward and horizontal solidification *vs.* upward solidification. For downward and horizontal solidification, ice dendrites show symmetric secondary dendritic arms (both sides of lamellae). By contrast, the misaligned dendrites show asymmetric secondary dendritic features in samples solidified vertically upward.

Ice lens defects, observed in the sintered $TiO_2$ walls as cracks oriented perpendicular to the direction of freezing, were observed in all samples solidified vertically upward for all initial particle fractions. None of the samples solidified vertically downward nor horizontally exhibit these defects. Two ice lensing regimes are observed: (i) ice spear defects (microcracks no wider than individual walls) were observed for samples fabricated using an initial particle fraction of 30 and 40 wt.% TiO$_2$, and (ii) ice banding for 50 wt.% TiO$_2$ samples, which corresponds to a merging of wall structures which increases over the height of the samples. In the last-to-solidify regions of these samples, a polygonal segregation pattern is present, which is consistent with a transition from ice-spear to periodic banding regimes described in the ice lensing literature.

Mean pore width and wall width are significantly smaller in samples solidified downward and horizontally in comparison to those solidified upward. This effect is greatest at 30 wt.% TiO$_2$, where mean pore width is more than four times larger when solidified upward compared to downward and approximately three times as large in comparison to horizontal solidification. Radial macrosegregation is observed for samples solidified upward (at a given height along the solidification direction, pore and wall width at the center of samples are larger than those near the walls of the mold) and is consistent with the development of a concave interface during solidification. Radial macrosegregation is attributed to a higher solidification velocity near the mold wall in comparison to that in the center of the samples and is not observed for downward and horizontal solidification configurations. For upward solidification, the effect is greatest at 30 wt.% TiO$_2$, where mean central and radial pore widths are 39±27 and 19±12 μm, respectively (corresponding to mean central and radial wall widths of 48±37 and 26±16 μm, respectively).

**Acknowledgements**
This work was supported by grants from NASA Office of Education and the Science Mission Directorate (NNH15ZDA010C), NASA's Physical Sciences Research Program (80NSSC18K0198), the Institute for Sustainability and Energy at Northwestern, and Northwestern University (NU) Office of the Provost. This work made use of the MatCI Facility and the EPIC facility (NUANCE Center) which are supported by the MRSEC program of the National Science Foundation (DMR-1121262) at the Materials Research Center at NU. The authors thank the following students for their assistance with experiments and ceramography: Mr. Youwu Fang (NU), Mr. Benjamin Richards (NU), and Ms. Catalina Young (NU, now at Purdue University). They also thank Dr. Christoph Kenel (NU) and Mr. Stephen Wilke (NU) for numerous useful discussions, and Prof. Peter Voorhees (NU) for helpful insights on radial macrosegregation in metal alloys.